\documentclass[12pt]{iopart}

\usepackage{iopams}  
\usepackage{graphicx}

\def\vq{{\bf q}}

\def\vk{{\bf k}}
\def\vQ{{\bf Q}}

\newcommand{\eq}[1]{(\ref{#1})}
\newcommand{\fig}[1]{figure~\ref{#1}}
\newcommand{\be}{\begin{equation}}
\newcommand{\ee}{\end{equation}}
\newcommand{\bea}{\begin{eqnarray}}

\newcommand{\eea}{\end{eqnarray}}

\begin{document}

\title{Spin nematic fluctuations near a spin-density-wave phase}

\author{Hiroyuki Yamase$^{1,2}$ and Roland Zeyher$^{1}$} 

\address{$^{1}$Max-Planck-Institut f\"ur Festk\"orperforschung,
             Heisenbergstrasse 1, D-70569 Stuttgart, Germany \\
$^{2}$National Institute for Materials Science, Tsukuba 305-0047, Japan}
\vspace{10pt}
\begin{indented}
\item 26 March 2015
\end{indented}

\begin{abstract}
We study an interacting electronic system exhibiting a spin nematic instability.
Using a phenomenological form for the spin fluctuation spectrum near the 
spin-density-wave (SDW) phase, we compute the spin nematic susceptibility in 
energy and momentum space as a function of temperature and the magnetic correlation length $\xi$. 
The spin nematic instability occurs when $\xi$ reaches a critical value $\xi_{\rm cr}$, 
i.e., its transition temperature $T_{\rm SN}$ 
is always higher than the SDW critical temperature  $T_{\rm SDW}$. 
In particular, $\xi_{\rm cr}$ decreases 
monotonically with increasing $T_{\rm SN}$. 
Concomitantly, low-energy nematic fluctuations are present in a wider 
temperature region as $T_{\rm SN}$ becomes higher. 
Approaching the spin nematic instability, the nematic spectral 
function at zero momentum exhibits a central peak as a function of energy for 
a finite temperature and a soft mode at zero temperature. 
These properties originate from the general 
feature that the imaginary part of the spin-fluctuation bubble has a term 
linear in energy 
and its coefficient is proportional to the square of temperature. 
Furthermore we find that the nematic spectral function exhibits a diffusive 
peak around zero momentum and zero energy without clear dispersive features.  
A possible phase diagram for the spin nematic and SDW transitions is also discussed. 
\end{abstract}

\pacs{74.70.Xa, 71.10.-w, 75.30.Fv, 75.40.-s} 
%
\vspace{2pc}
\noindent{\it Keywords}: iron-based superconductors, spin nematic fluctuations, spin-density-wave 
%
%
%
%

\section{Introduction}
The mechanism of high-temperature superconductivity is one of the major
issues in condensed matter physics. By carrier doping into antiferromagnetic 
Mott insulators, cuprate superconductors attain high critical temperatures 
$T_c$, typically of more than 77 K, 
the boiling temperature of liquid nitrogen at ambient pressure.
Since it is widely accepted that doped cuprates can be described within a
one-band model it is natural to assume that the important fluctuations 
are spin fluctuations in these systems. 
In 2008 another family of high-$T_c$ superconductors was discovered consisting of
iron-based pnictides \cite{kamihara08}. Similar to the cuprate case, 
superconductivity occurs close to a magnetic phase. 
However, in contrast to the cuprates, the magnetic phase is metallic and 
several bands cross the Fermi energy, suggesting 
the importance of orbital degrees of freedom. Thus there are two different
kinds of fluctuations in the pnictides which may be relevant for
their physical properties, namely, spin and orbital fluctuations. 

In general it is not easy to assess the relative importance of the two kinds
of fluctuations in observable quantities. It is known that superconducting
order parameters due to spin fluctuations tend to have $s_{\pm}$ symmetry
\cite{mazin08,kuroki08} 
whereas those due to orbital fluctuations have    
$s_{++}$ symmetry \cite{yanagi10,kontani11}. While this 
difference in the gap symmetries can contribute   
to identifying the underlying mechanism of superconductivity, 
a recent theoretical study \cite{yamada14} shows that $s_{\pm}$ pairing gap 
can be stabilized even for orbital fluctuations when a partial 
contribution from spin fluctuations is taken into account. 
Experimentally the superconducting phase occurs often closer to the 
nematic than to the magnetic phase \cite{stewart11,fisher11,kasahara12}.
From this one may conclude that nematic fluctuations 
play also an important role for superconductivity and more generally for the
physics of pnictides. It is thus important to study the properties of 
nematic fluctuations and their origin in detail. 

The nematic instability is accompanied by a 
structural phase transition from a tetragonal to an orthorhombic phase. 
This structural phase transition is believed to be driven by the coupling 
to electronic degrees of freedom \cite{fisher11}, because the observed 
anisotropy of 
resistivity \cite{chu10}, the optical conductivity \cite{nakajima11} and  
the splitting of electronic bands \cite{yi11} is much larger than what one 
would expect from the lattice anisotropy. Due to these effects the orthorhombic 
phase in the pnictides, which breaks the $C_4$ symmetry of the normal 
state, is referred to as an electronic nematic phase. 
Depending on the electronic degrees of freedom responsible for the nematicity, 
three kinds of nematicity can be distinguished: charge 
\cite{kivelson98,yamase00a,yamase00b,metzner00}, 
orbital \cite{raghu09,wclee09}, and spin nematicity \cite{andreev84}. 
For iron-based pnictides, the latter two possibilities seem to be relevant. 
 
Orbital nematicity in pnictides is associated with a spontaneous difference 
in the occupation of $d_{yz}$ and $d_{zx}$ orbitals 
\cite{krueger09,cclee09,lv09}. Below the transition temperature the  
orientational symmetry is broken, whereas all the other symmetries 
remain unbroken. Orbital fluctuations  may lead to a 
superconducting state driven mainly by fluctuations around zero momentum 
\cite{yanagi10}. In this scenario, nesting of the Fermi surfaces 
is not crucial for pairing. Orbital fluctuations 
around a finite momentum \cite{stanescu08,kontani11} may, similarly to 
spin fluctuations \cite{mazin08,kuroki08}, also lead to superconductivity
but in this case nesting properties of the Fermi surface are important.
Considering general orbital fluctuations, retardation effects 
and quasi-particle weights within Eliashberg theory it was shown  \cite{yamase13b} that 
orbital nematic fluctuations produce strong coupling 
superconductivity. The obtained transition temperatures 
and the fact that superconductivity also occurs inside the nematic
phase are compatible with the experimental observations.

Spin nematicity in iron-based pnictides is associated with a spontaneous 
anisotropy of the spin fluctuation spectrum in momentum space 
between $(\pi,0)$ and $(0,\pi)$ \cite{fang08,xu08}. Spin nematicity owes its
existence to spin fluctuations and is caused by the interaction between
them as reviewed in \cite{fernandes12} and \cite{fernandes14}. It
has the general property that it can occur 
above the spin-density-wave (SDW) phase \cite{fernandes12}, which is in nice
agreement with the experimental findings. 

In general it is difficult to distinguish between spin and orbital
nematicity because there is a coupling between the corresponding
order parameters. If the orbital (spin) nematic instability occurs, 
the spin (orbital) nematic order is induced via a linear coupling between 
them \cite{fernandes14}. Hence both orbital and spin 
nematic orders occur always at the same time. Nevertheless, one may
argue that the original instability occurs only in the orbital or the 
spin section. As a result one mechanism will dominate 
and it makes sense to study the two kinds of mechanisms leading to the nematic
state independently. 

Using linear response theory we will investigate in this paper properties 
near the spin nematic instability. The results are rather independent of
microscopic details because the functional form for the spin fluctuation
propagator becomes generic close to the SDW phase. The
characteristic features of the low-energy spin nematic fluctuations are 
determined by the anomalous asymptotic behavior of the spin fluctuation 
bubble at low frequencies. Our results for the dynamic structure factor are
predictions which can be checked by inelastic light scattering. 
We also present a schematic phase diagram containing spin nematic and magnetic
transition temperatures.

\section{Model and formalism}
The spin nematicity in iron-based superconductors can be formulated in terms of 
an action as in \cite{fernandes12a}. 
Here we wish to formulate it in terms of a conventional operator formalism 
and to clarify the diagrammatic structure of spin nematic physics. 

Iron-based superconductors are often described by a five-band Hubbard model \cite{kuroki08}. 
To describe the spin nematic interaction in a microscopic model, 
we focus on the effective interaction of the total spin operator. 
Then our microscopic model consists of electrons hopping between the sites of a
square lattice and interacting via their total spins. The corresponding
Hamiltonian becomes 
\be
H = \sum_{{\bf k},\alpha,\sigma} \epsilon_{{\bf k}\alpha} 
c^\dagger_{{\bf k},\alpha,\sigma} c_{{\bf k},\alpha,\sigma} +\frac{1}{2N}
\sum_{\bf q} J({\bf q})\; {\bf S}({\bf q}) \cdot {\bf S}(-{\bf q}),
\label{H}
\ee
with the total spin operator
\be
{\bf S}({\bf q}) = \frac{1}{2} \sum_{{\bf k},\alpha,\sigma,\sigma'} 
c^\dagger_{{\bf k},\alpha,\sigma} {\boldsymbol \tau}_{\sigma \sigma'}
c_{{\bf k}+\vq,\alpha,\sigma'}.
\label{S}
\ee
$c^\dagger_{{\bf k},\alpha,\sigma}$ creates an electron in the band $\alpha$ 
with spin direction 
$\sigma$, and $c_{{\bf k},\alpha,\sigma}$ annihilates this electron. {\boldmath $\tau$}
is the vector of the three Pauli matrices, $\epsilon_{{\bf k}\alpha}$ the 
one-particle energies of the band $\alpha$, $J({\bf q})$ the strength
of the spin-spin interaction in momentum space, and $N$ is the total number of 
the lattice sites. 

Besides the total spin operator we will consider the spin 
nematic operator
\be
\Phi({\bf q}) = \frac{1}{\sqrt{2}N} \sum_{\bf k} \gamma_{\bf k}\; 
{\bf S}({\bf k}+{\bf q}/2)\cdot {\bf S}({\bf -k}+{\bf q}/2).
\label{Phi}
\ee
$\gamma_{\bf k}$ is equal to one if ${\bf k}$ is close to $(\pi,0)$ or
$(-\pi,0)$ and equal to minus one if ${\bf k}$ is close to $(0,\pi)$ or
$(0,-\pi)$. $\gamma_{\bf k}$ has $d$-wave symmetry and the sum over ${\bf k}$ in 
Eq. (\ref{Phi}) runs over the entire Brillouin zone (BZ).

The dynamic spin susceptibility is defined by
\be
\chi({\bf q},\omega) = \frac{i}{N} \int_0^\infty dt \langle 
[{\bf S}_l({\bf q},t),{\bf S}_l(-{\bf q},0)] \rangle e^{i\omega t},
\label{chi}
\ee
where the index $l$ denotes a Cartesian component, $[\cdot , \cdot]$ the commutator, 
and $\omega$ includes tacitly a small imaginary part $i \eta$; we consider a state with 
spin rotational symmetry and thus $\chi$ does not depend on $l$. 
In a similar way we define a spin nematic susceptibility by
\be
\chi_{\rm SN}({\bf q},\omega) = \frac{i}{N} \int_0^\infty dt \langle 
[\Phi({\bf q},t),\Phi(-{\bf q},0)] \rangle e^{i\omega t}.
\label{chinem}
\ee

Going over to the Matsubara representation, the usual diagrammatic perturbation
expansion for electronic systems holds for the above two correlation functions.
Concerning $\chi({\bf q},\omega)$ we are interested in the following in its
low-frequency and long-wavelength behavior. Here long-wavelength means
momenta near $(\pi,0)$ or $(0,\pi)$ or the equivalent points in the BZ. Thus
we do not try to calculate $\chi$ by a perturbation expansion but expand the
inverse susceptibility in powers of $\omega$ and ${\bf q}-{\bf Q}_{\bf q}$
where ${\bf Q}_{\bf q}$ denotes $(\pi,0)$ for ${\bf q}$ close to 
$(\pi,0)$ and $(0,\pi)$ for ${\bf q}$ close to $(0,\pi)$. The result is 
\be
\chi({\bf q},\omega) = \frac{c/\gamma}{r+({\bf q}-{\bf Q}_{\bf q})^2 
-i\omega/\gamma}.
\label{chi-1}
\ee
$r$ is equal to $1/\xi^2$ where $\xi$ is the correlation length. $\gamma$
is a damping constant and $c$ determines the spectral weight. The above
parametrization of $\chi$ is a general form of spin fluctuations near the SDW phase 
and in fact describes rather well the measured imaginary part
of the spin susceptibility in BaFe$_{1.85}$Co$_{0.15}$As$_2$ by choosing 
appropriate parameters \cite{inosov10}.

\begin{figure}  [ht]
\vspace*{0cm}
\begin{center}
\includegraphics[angle=0,width=11.0cm]{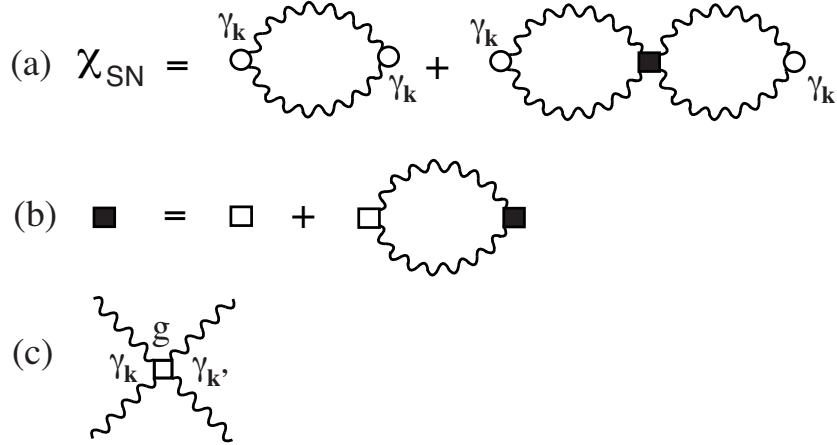} 
\end{center}
\caption{Graphical representation of the spin nematic susceptibility $\chi_{\rm SN}$. 
The wavy line indicates spin fluctuations 
and the vertex with a circle in (a) represents a form factor $\gamma_{\vk}$ associated with 
the spin nematic order parameter [see \eq{Phi}]. The solid square is  an 
effective four-point vertex of spin fluctuations defined in (b). 
The open square in (b) and (c) is a bare four-point vertex of spin fluctuations 
due to the spin nematic interaction [see \eq{HI}]. 
}
\label{diagram} 
\end{figure}

The lowest order diagrams for the evaluation of $\chi_{\rm SN}$ are shown
in \fig{diagram}(a). The wavy line stands for the spin propagator $-\chi$, i.e., for
a two-particle propagator. The first diagram represents $\chi^{(0)}_{\rm SN}$
for the non-interacting case, whereas the second term comes from the spin nematic 
interaction. The effective interaction (solid square) represents  
the four-spin vertex as shown in figures~\ref{diagram}(b) and (c). 
The open square denotes a ring
diagram where the sides of the square represent electronic Green's functions
and the corner points of the square the interaction term in Eq. (\ref{H}).
Taking the low-frequency, long-wavelength limit in external spin legs
the four-spin vertex collapses to one constant $g$. The effective interaction 
term assumes then the form
\be
H_I = -\frac{g}{2N} \sum_{\bf q} \Phi({\bf q}) \Phi(-{\bf q}).
\label{HI}
\ee
The resulting diagrams for $\chi_{\rm SN}$ are shown in Fig. 1. Altogether
one obtains
\be
\chi_{\rm SN}({\bf q},\omega) = \frac{\chi^{(0)}_{\rm SN}({\bf q},\omega)}
{1-g \chi^{(0)}_{\rm SN}({\bf q},\omega)}.
\label{RPA}
\ee
The analytic expression for $\chi^{(0)}_{\rm SN}$ is
\be
\chi^{(0)}_{\rm SN}({\vq},i\omega_n) = \frac{3T}{N} \sum_{\vk,m} \chi({\bf k}+{\bf q}/2,
i\omega_{n}+ i\nu_m) \chi(-{\bf k} +{\bf q}/2, -i\nu_m),
\label{chi0}
\ee
where $\omega_n$ and $\nu_m$ are bosonic Matsubara frequencies, 
the factor 3 comes from spin rotational symmetry and $T$ is the temperature. 
In the long-wavelength and low-frequency limit follows from the above equation
for small $r$,
\be
\chi^{(0)}_{\rm SN}({\bf 0},0) = \frac{3T}{N}\sum_{\vk, m} \left[\chi(\vk,i\nu_m)\right]^2 >
\frac{3T}{N} \sum_{\vk} \left[ (\chi(\vk,0)\right]^{2} \sim \frac{T}{r}.
\label{inequal}
\ee
Assuming $g$ to be positive [attractive interaction in \eq{HI}], 
\eq{inequal} implies that $\chi_{\rm SN}({\bf 0},0)$
will diverge at a finite temperature for a finite and positive $r$
independently how large the coupling constant $g$ is. This means that a
nematic transition will always take place before the long-range 
magnetic phase at $r=0$ is reached if $g$ is positive and $T$ finite. 

In order to evaluate Eq. (\ref{chi0}) we use the spectral representation
for $\chi$,
\be
\chi({\bf q},i\omega_n) = \int_{-\infty}^{\infty} d \epsilon \; \frac{A({\bf q},\epsilon)}{i\omega_n -
\epsilon},
\label{spectral}
\ee
with 
\be
A({\bf q},\epsilon) = -\frac{c}{\pi\gamma} \frac{\epsilon/\gamma}
{\left[ r+(\vq - \vQ_{\vq})^2\right]^2 + \epsilon^2/\gamma^2}.
\label{A}
\ee
Inserting \eq{spectral} into \eq{chi0} and carrying out the
frequency sum yields after an analytic continuation for the imaginary part
\be
\hspace{-20mm} {\rm Im} \;\chi^{(0)}_{\rm SN}({\vq},\omega) = \frac{3\pi}{N} \sum_{\vk} 
\int_{-\infty}^{\infty} d \epsilon A(\vk+\vq/2,\epsilon+\omega)
A(-{\bf k}+{\bf q}/2,\epsilon) 
\left[ {n(\epsilon)-n(\omega + \epsilon)} \right],
\label{Ichi}
\ee
with $n(\epsilon) = 1/(e^{\epsilon/T}-1)$.  
The real part of $\chi^{(0)}_{\rm SN}$ is computed via a Kramers-Kronig relation, 
\be
{\rm Re} \chi^{(0)}_{\rm SN}(\vq,\omega)=\frac{1}{\pi}{\rm p.v.} 
\int_{-\infty}^{\infty} d\nu \frac{{\rm Im} \chi^{(0)}_{\rm SN}({\vq},\nu)}{\nu-\omega} \,,
\label{Rchi}
\ee
where p.v. denotes the principal value. The function 
$\chi_{\rm SN}({\bf q},\omega)$ 
is then obtained from (\ref{RPA}), (\ref{Ichi}) and (\ref{Rchi}).    
The spectral function of the spin nematic fluctuations is given by
\be
S(\vq,\omega)=\frac{1}{\pi}\left[1+n(\omega)\right]{\rm Im}\chi_{\rm SN}(\vq,\omega)\,,
\label{sqw}
\ee
which we compute numerically. 

\section{Results}
\subsection{Choice of parameters} 
To compute the spin nematic spectrum numerically from (\ref{RPA}) 
and (\ref{chi0}),  
we first fix the spin fluctuation propagator [see \eq{chi-1}] and the coupling 
strength $g$ [see \eq{HI}]. 
Neutron scattering data \cite{inosov10} 
for BaFe$_{1.85}$Co$_{0.15}$As$_{2}$ yield parameters for the spin fluctuation 
spectrum, namely,  $c\approx 1.3$ and $\gamma \approx 230$ meV.
Hence we take in the present theory
\be
c=1 \quad {\rm and} \quad \gamma=1\,,
\label{c-gamma}
\ee 
measuring all quantities with the dimension of energy in units of $\gamma$. 
For the mass term $r$ we take in the normal state 
\be
r = r_{\rm cr}+ (T-T_{\rm SN}) \quad {\rm for} \quad T \geq T_{\rm SN}.
\label{r}
\ee
The bare susceptibility $\chi^{(0)}_{\rm SN}({\bf 0},0)$ depends on $T$
and $r$. The instability equation of the normal state with respect
to the nematic state [see \eq{RPA}],
\be
1 = g \chi^{(0)}_{\rm SN}({\bf 0},0),
\label{instabil}
\ee
yields a relation between the critical mass $r_{\rm cr} = r(T_{\rm SN})$ and the 
nematic transition temperature $T_{\rm SN}$. 
The value of $T_{\rm SN}$ (or equivalently $r_{\rm cr}$) can be considered 
as a free parameter and we will study 
the spin nematic spectrum for various choices of $T_{\rm SN}$. 
At $T=0$,  \eq{r} is not valid and $r$ should be taken as a 
non-thermal control parameter, for example, concentration of an isovalent 
substitution, carrier density, or pressure. 

According to the phase diagram of Ba(Fe$_{1-x}$Co$_{x}$)$_{2}$As$_{2}$ in 
\cite{nandi10}, a structural phase transition and the SDW instability occur 
at 80 K and 70 K, respectively, at $x=0.04$. 
If the structural phase transition is assumed to be associated with the spin 
nematic instability, the present theory reproduces the experimental transition 
temperatures at 
$x=0.04$ for $g=0.29$, extrapolating $r$ linearly down to $T_{\rm SDW}$. 
Since a different value of $g$ would be obtained if one considers experimental data 
at a different $x$, we will also take $g=0.9$ later to clarify how our results depend on
the choice of $g$.

We first study the case of $g=0.29$, 
and calculate the critical value $r_{\rm cr}$ as a function of $T_{\rm SN}$ 
(\fig{TSN-rcr}) and the   
spectral weight of spin nematic fluctuations in $\vq$ and $\omega$ space 
(figures~\ref{S00-map}, \ref{S0w-w}, and \ref{Sqw-map}). 
We will then consider results for a larger coupling strength $g=0.9$ 
(figures~\ref{TSN-rcr}, \ref{S0w-w-soft}, and \ref{S0w-w-cross}). 
On the basis of these results, we sketch typical phase diagrams of the spin nematic 
phase near the SDW phase (\fig{T-delta}). 

\subsection{Numerical results} 

\begin{figure}  [ht]
\vspace*{0cm}
\begin{center}
\includegraphics[angle=0,width=8.0cm]{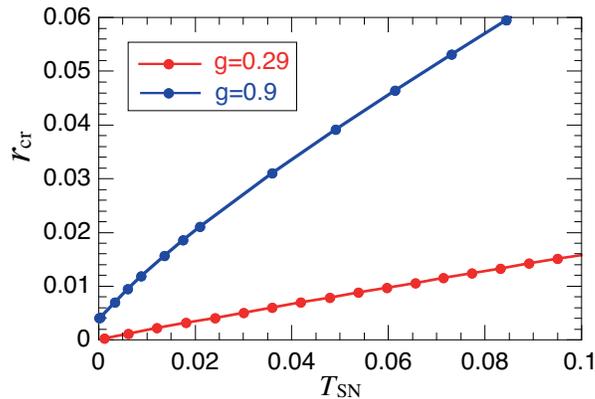} 
\end{center}
\caption{\label{TSN-rcr} 
(Color online) 
Critical value $r_{\rm cr}$ as a function of the
onset temperature of the spin nematic instability $T_{\rm SN}$ for 
a spin nematic coupling strength of $g=0.29$ and $0.9$. 
}
\end{figure}

We put $\vq \rightarrow {\bf 0}$ and $\omega \rightarrow 0$ in \eq{RPA} and 
study the spin nematic instability.  We search for its onset 
temperature $T_{\rm SN}$ by determining the temperature at which 
$\chi_{\rm SN}({\bf 0},0)$ diverges for a given $r_{\rm cr}$. Figure~\ref{TSN-rcr}  
shows $r_{\rm cr}$ as a function of $T_{\rm SN}$ for two coupling strengths $g$.
$r_{\rm cr}$ is always positive and increases with increasing  $T_{\rm SN}$. 
This feature can easily be understood by using 
the approximate expression \eq{inequal} for  $\chi^{(0)}_{\rm SN}$ which holds 
at large temperatures. The instability condition Eq. (\ref{instabil})
becomes then equivalent to $r_{\rm cr} \sim  g T_{\rm SN}$.
The linear dependence of $r_{\rm cr}$ with $T_{\rm SN}$ is at least approximately 
reflected in the calculated curves in \fig{TSN-rcr}. The same holds for the 
increase of the slope with increasing $g$.

\begin{figure}[ht]  
\vspace*{0cm}
\begin{center}
\includegraphics[angle=0,width=10cm]{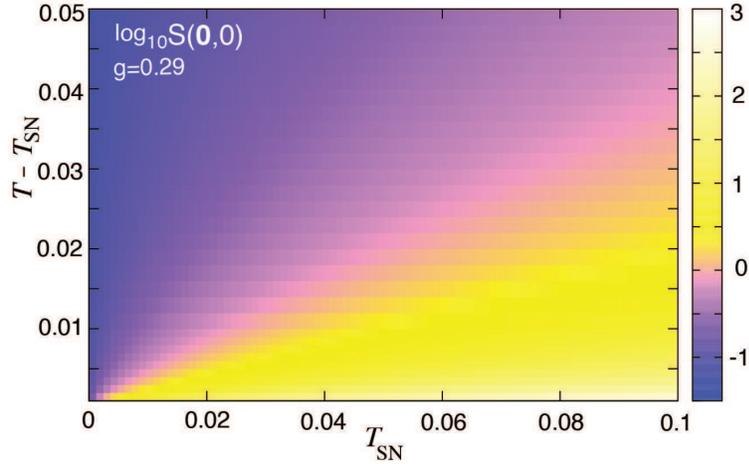} 
\end{center}
\caption{\label{S00-map}
(Color online) 
Spectral weight of the spin nematic fluctuations at $\vq={\bf 0}$ and $\omega=0$ 
for $g=0.29$ in the plane of $T_{\rm SN}$ and $T-T_{\rm SN}$; 
the spectral weight is plotted on a logarithmic scale. 
}
\end{figure}

Figure~\ref{S00-map} shows the spectral weight of the spin nematic fluctuations 
at $\vq={\bf 0}$ and $\omega=0$ in the plane of $T_{\rm SN}$ and $T-T_{\rm SN}$. 
The spectral weight is enhanced upon approaching $T_{\rm SN}$ 
and eventually diverges at $T_{\rm SN}$. 
In particular, strong fluctuations appear at higher temperatures 
above $T_{\rm SN}$ 
if $T_{\rm SN}$ becomes larger, as seen in the 
yellow region in \fig{S00-map}. For $T_{\rm SN} \approx 0$,  on the other hand, 
the enhancement of the spectral weight occurs only very close to 
$T=T_{\rm SN}$.  

\begin{figure} [ht]
\vspace*{0cm}
\begin{center}
\includegraphics[angle=0,width=8.0cm]{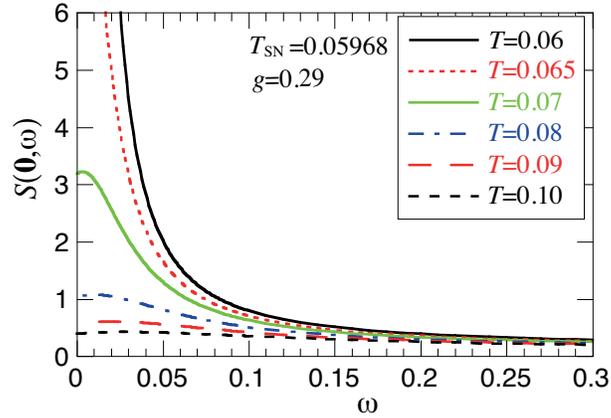} 
\end{center}
\caption{\label{S0w-w}
(Color online) 
$\omega$ dependence of the spectral weight at $\vq={\bf 0}$ 
for $g=0.29$ and different temperatures. The spin nematic instability 
occurs at $T_{\rm SN}=0.05968$ with the critical value $r_{cr} = 0.00968$.
}
\end{figure}

Retaining still $\vq={\bf 0}$, we next present in \fig{S0w-w} 
the dependence of the spectral weight on $\omega$  
for several temperatures. At high temperature well above $T_{\rm SN}$ 
the spectrum is almost flat at low energies and its weight is very small. 
With decreasing temperature the low-energy spectral weight is enhanced 
to form a peak at zero energy in form of a central peak. 
This peak grows more and more upon approaching $T_{\rm SN}$ 
and finally diverges at $T_{\rm SN}$.

\begin{figure}  [ht]
\vspace*{0cm}
\begin{center}
\includegraphics[angle=0,width=10.0cm]{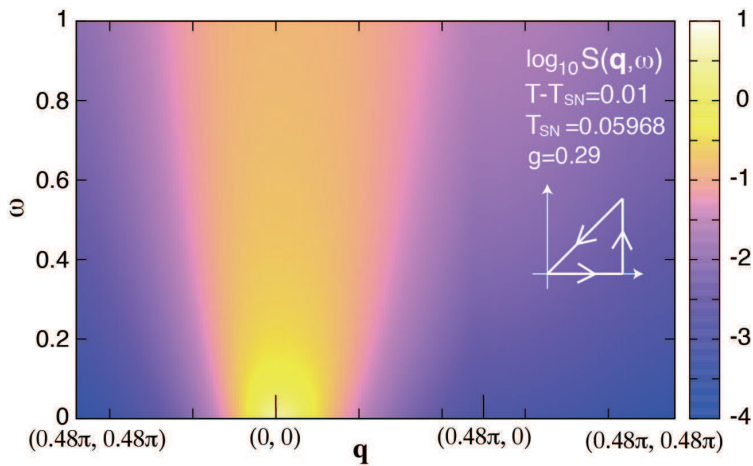} 
\end{center}
\caption{\label{Sqw-map}
(Color online) 
$\vq$-$\omega$ map of the spectral weight on a logarithmic scale 
for $g = 0.29$. 
The scanned direction of $\vq$ is sketched inside the figure: 
$(0.48\pi, 0.48\pi)$ $\rightarrow$ $(0,0)$ $\rightarrow$ $(0.48\pi, 0)$ 
$\rightarrow$ $(0.48\pi, 0.48\pi)$. The spectrum has a fourfold symmetry in 
momentum space above $T_{\rm SN}$.  
}
\end{figure}

Figure~\ref{Sqw-map} is a $\vq$-$\omega$ map of the spectral weight 
near the spin nematic instability. The highest spectral weight is located 
around $\vq={\bf 0}$ and $\omega=0$ and the weight spreads with increasing 
energy while its strength is decreasing, like the tail of a comet. 
That is, the spin nematic fluctuations appear as a diffusive peak around 
$\vq={\bf 0}$ and $\omega=0$, and no dispersive features can be seen. 
While we have chosen $T_{\rm SN}=0.05968$ and $T-T_{\rm SN}=0.01$ 
in \fig{Sqw-map}, 
essentially the same result is obtained for other choices of parameters, 
although the {\it tail of the comet} is gradually blurred with increasing $T$.

We next comment on results at $T_{\rm SN}=0$. 
Figure~2 shows that for the coupling strength $g=0.29$ $r_{\rm cr}$ 
almost vanishes at $T_{\rm SN}=0$. 
Hence the spin nematic and SDW instability occur almost simultaneously. 
Collective effects of spin nematic 
fluctuations occur only in the vicinity of  $T=T_{\rm SN}$ 
as can be inferred from \fig{S00-map}.   
As a result spin nematic fluctuations are enhanced only well 
below $T\approx 10^{-4}$ for $T_{\rm SN} = 0$.  
If a larger value for the coupling strength $g$ is used, 
spin nematic fluctuations for $T_{\rm SN}=0$ become 
visible at much higher temperatures. 
For $g=0.9$, for instance,    
the overall temperature scale associated 
with the spin nematic instability becomes much larger than for $g = 0.29$ 
and low-energy fluctuations are strongly enhanced already below 
$T \approx 0.02$ [$\approx 50$ K, see \eq{c-gamma} and also \fig{S0w-w-cross}].
We have checked also that there is no qualitative change in  figures~\ref{S00-map}, \ref{S0w-w}, 
and \ref{Sqw-map} if a finite $T_{\rm SN}$ is considered and $g$ is increased from 0.29
to 0.9. 

\begin{figure}  [ht]
\vspace*{0cm}
\begin{center}
\includegraphics[angle=0,width=8.0cm]{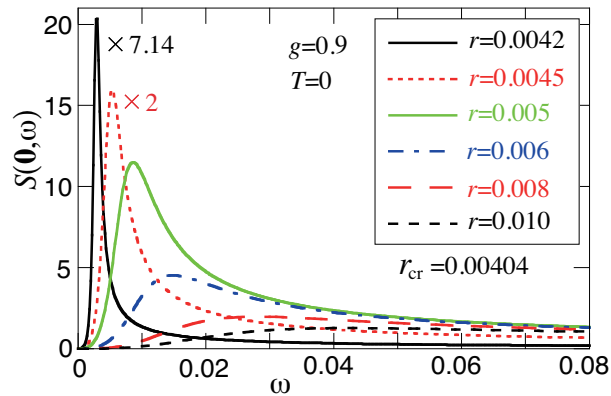} 
\end{center}
\caption{\label{S0w-w-soft}
(Color online) $\omega$ dependence of the spin nematic spectral weight  
at $\vq={\bf 0}$ for several values of $r$.  
The weights for $r=0.0045$ and $0.0042$ are actually larger by a factor 2 and 7.14,
respectively, as indicated in the figure. 
}
\end{figure}

At zero temperature we consider $r$ in \eq{chi-1} as a non-thermal control parameter, 
see also the statement below \eq{instabil}. We plot in \fig{S0w-w-soft} the $\omega$ dependence of 
the spin nematic spectral weight at $\vq={\bf 0}$ for several values of $r$ 
above the spin nematic instability at $r_{\rm cr}=0.00404$ for $g=0.9$. 
In contrast to the case of a finite $T$, described in \fig{S0w-w}, 
no central peak is formed and the weight at $\omega=0$ remains zero 
when approaching the spin nematic instability. Instead spin nematic fluctuations 
appear as a soft mode. With decreasing $r$  a peak structure forms at a finite energy
and moves towards lower frequencies. When  $r$ approaches $r_{\rm cr}$, the width of the
peak becomes very narrow and at the same time the height of the peak increases strongly. 
At the spin nematic transition $r=r_{\rm cr}$ the weight diverges at $\omega=0$.  
The presence of the soft mode suggests that there might be a dispersive feature of the 
spin nematic fluctuations in the plane of $\vq$ and $\omega$ at zero temperature. 
Hence we computed maps of the spectral weight, similar to \fig{Sqw-map}, at $T=0$. 
The dispersive feature is actually obtained, but only for $r$ very close to $r_{\rm cr}$; 
furthermore it becomes visible only in the vicinity of $\vq={\bf 0}$ and $\omega=0$, 
on a much smaller scale than that in \fig{Sqw-map}. 
In this sense, the dispersive feature is extremely weak. 
In fact, once $r$ becomes a little larger than $r_{\rm cr}$,  
spin nematic fluctuations produce a diffusive signal 
around $\vq={\bf 0}$ and $\omega=0$ even at $T=0$, similar as in \fig{Sqw-map}.

\subsection{Asymptotic behavior of $\chi^{(0)}_{\rm SN}$ at low frequencies}
 
Our obtained results can easily be understood by analyzing 
the low-energy property of the spin-fluctuation bubble 
$\chi^{(0)}_{\rm SN} ({\bf q},\omega)$. 
From \eq{Ichi} follows ${\rm Im} \chi^{(0)}_{\rm SN}(\vq,-\omega) = 
- {\rm Im}\chi^{(0)}_{\rm SN}({\bf q},\omega)$, i.e., 
${\rm Im}\chi^{(0)}_{\rm SN}({\bf q},\omega)$ is
an odd function of the frequency. 
Evaluating \eq{Ichi} for $\omega, T \ll r$ yields the asymptotic expansion 
\be
{\rm Im}\chi_{\rm SN}^{(0)}({\bf 0},\omega) \propto 
\frac{T^2}{r^3}\omega + a_3 \; \omega^3 + \cdots \,,
\label{asymp}
\ee
where $a_3$ is a constant. 
There is a term linear in $\omega$, which yields a central peak as found in 
\fig{S0w-w}. Its coefficient has a $T^{2}$ dependence, leading in  \fig{S00-map}
to strong fluctuations over a wide temperature region above $T_{\rm SN}$ for a higher 
$T_{\rm SN}$. On the other hand, at $T=0$, the liner term 
vanishes and the spectral weight is characterized by $\omega^{3}$ at low 
$\omega$, leading to a substantial suppression of the spin nematic 
fluctuations at low $\omega$. This is the reason why spin nematic fluctuations 
are strongly suppressed at low temperatures (\fig{S00-map}) and no central peak is present. 
Instead a soft mode associated with spin nematic fluctuations occurs 
at $T=0$ as shown in \fig{S0w-w-soft}. Since the spin-fluctuation propagator 
is in general parameterized by \eq{chi-1} close to the SDW 
instability, the low-energy dependence of \eq{asymp} is understood as a general 
feature of the spin-fluctuation bubble. This low-energy property yields 
characteristic features of the spin nematic fluctuations found in 
figures~\ref{S00-map} - \ref{S0w-w-soft} as well as the spin nematic 
instability close to the SDW phase shown in \fig{TSN-rcr}. 

\begin{figure}  [ht]
\vspace*{0cm}
\begin{center}
\includegraphics[angle=0,width=8.0cm]{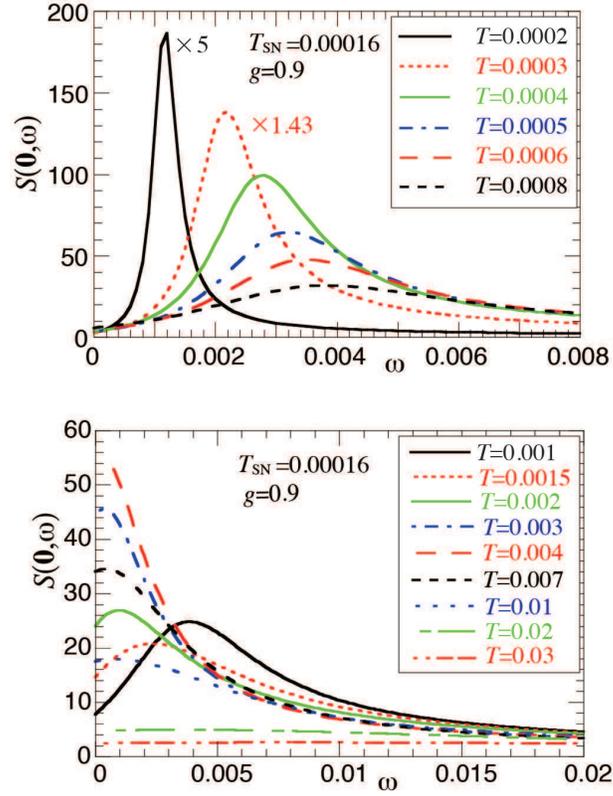} 
\end{center}
\caption{\label{S0w-w-cross}
(Color online) 
Evolution of the spectral weight for $g=0.9$ and $T_{\rm SN}=0.00016$ 
as a function of temperature. The critical value $r_{\rm cr}$ is 0.00406 (see \fig{TSN-rcr}). 
}
\end{figure}

The peculiar behavior of $\chi^{(0)}_{\rm SN}({\bf 0},\omega)$ can be seen most 
clearly in \fig{S0w-w-cross}. 
The two diagrams in this figure show the evolution
of the spectral weight for $g=0.9$ as a function of temperature. $T_{\rm SN}$
is equal to 0.00016 
and thus very small, yielding spectra which are also
representative for $T_{\rm SN} = 0$. 
Fixing $T_{\rm SN}$ means that also $r_{\rm cr}$ 
is fixed, see \eq{r} and \fig{TSN-rcr}.  
The lower diagram shows that for 
$T \gtrsim 0.02$ the spectrum is
completely flat and structureless. Decreasing $T$ down to 0.004 
the spectral weight shifts towards lower frequencies and a central
peak is formed. In this temperature range the linear term
in $\omega$ of ${\rm Im} \chi^{(0)}_{\rm SN}({\bf 0},\omega)$ still dominates 
in a low frequency expansion of ${\rm Im} \chi^{(0)}_{\rm SN}$ 
[see \eq{asymp}] and causes the central peak. 
Lowering further $T$ to 0.001 the low frequency part of the spectral weight 
looses intensity and a propagating peak appears with an increasing frequency 
and a decreasing half-width. In this temperature interval the linear term
in $\omega$ of ${\rm Im} \chi^{(0)}_{\rm SN}({\bf 0},\omega)$, proportional to
$T^2$, becomes small and the term $a_3 \omega^3$ starts to play a role.
As a result a transition from a diffusive to a propagating mode behavior
is obtained.  Considering the upper diagram 
and decreasing further the temperature the energy of the propagating
peak and its half-width decrease whereas its height increases strongly.

\subsection{Schematic phase diagram}   
\begin{figure}  [ht]
\vspace*{0cm}
\begin{center}
\includegraphics[angle=0,width=8.0cm]{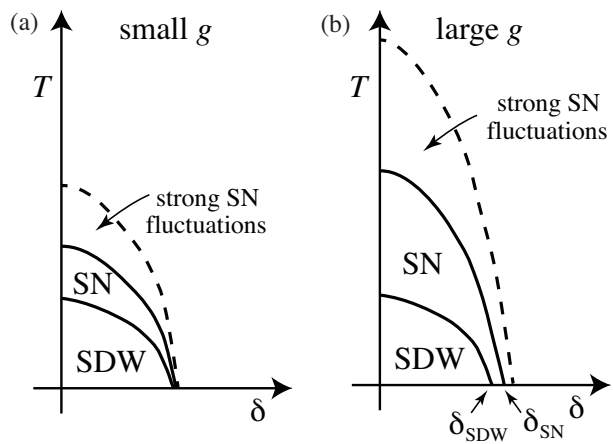} 
\end{center}
\caption{\label{T-delta} 
Schematic phase diagram of the spin nematic (SN) instability near the 
SDW phase on the plane of a control parameter $\delta$ and the temperature 
for a realistic value of $g$ for iron pnictides (left figure) and  for a 
large $g$ (right figure).
The phase transition is assumed to be continuous. Two quantum phase transition 
occur at $\delta_{\rm SN}$ and $\delta_{\rm SDW}$, and these two almost 
coincide for a small $g$. 
}
\end{figure}

On the basis of our obtained results (figures~\ref{TSN-rcr} and \ref{S00-map}), 
we can infer a typical phase diagram for the spin nematic and SDW
phases. Let the onset temperature of the SDW instability evolve as a function of 
a control parameter $\delta$, as shown in \fig{T-delta}(a); 
$\delta$ may correspond to the
concentration of substituted ions, carrier density, pressure, or other 
quantities depending on material properties. $T_{\rm SDW}$ decreases with 
increasing $\delta$ and vanishes at a critical value of $\delta_{\rm SDW}$. 
Let us introduce the slope $\alpha = r_{\rm cr}/(T_{\rm SN}-T_{\rm SDW})$ $(>0)$ 
and assume that $\alpha$ depends only weakly on $\delta$. 
Approximating the results in \fig{TSN-rcr} by $r_{\rm cr} \propto T_{\rm SN}$,  
we obtain $T_{\rm SN}-T_{\rm SDW} \propto T_{\rm  SDW}$, 
i.e., the temperature region occupied by the spin nematic phase increases 
with increasing $T_{\rm SDW}$.
In the opposite limit where $T_{\rm SDW}$ 
is close to zero, i.e., $\delta \rightarrow \delta_{\rm SDW}$, 
the spin nematic phase vanishes  very close to $\delta_{\rm SDW}$.   
As seen from \fig{S00-map}, the temperature region where strong spin 
nematic fluctuations are present expands  for a higher $T_{\rm SN}$ and 
shrinks substantially near $\delta_{\rm SDW}$ as shown by the dashed line in 
\fig{T-delta}(a). On the other hand, if the strength of the spin nematic 
interaction  becomes very strong, the spin nematic phase as well as the 
region where strong spin nematic fluctuations 
are present expands to a larger region as shown in \fig{T-delta}(b). In particular, 
a region of the spin nematic phase can be well separated from the SDW 
instability at zero temperature (see also \fig{TSN-rcr} for $g=0.9$) and 
thus two well-separated quantum phase transitions exist as a function of $\delta$. 

\section{Discussions}
\subsection{Origin of the nematic phase in iron-based superconductors} 
We discuss the origin of the nematic phase observed in iron-based superconductors \cite{fisher11}. 
In \fig{T-delta}, 
we assumed that $T_{\rm SDW}$ decreases monotonically with increasing $\delta$, 
which is the usual case in iron-based superconductors. 
The spin nematic scenario then predicts that 
the temperature difference of $T_{\rm SN}$ and $T_{\rm SDW}$, namely 
$r_{\rm cr}/\alpha = T_{\rm SN}-T_{\rm SDW} >0$, should decrease monotonically 
with increasing $\delta$, at least, if $\alpha$ depends only weakly
on $\delta$. 
However, this tendency is not observed in iron-based 
superconductors in spite of the fact that many different compounds 
\cite{stewart11} have been investigated. 
We discuss several possibilities to resolve this qualitative discrepancy between 
the expectation and the experiment.

First, we have assumed a constant coupling strength $g$ 
and a constant value of $c$; the latter controls the overall strength of spin fluctuations [see \eq{chi-1}]. 
If $g$ (and/or $c$) is substantially suppressed at a low $\delta$, $T_{\rm SN}$ 
would shift closer to $T_{\rm SDW}$ similar to the 
experimental observation. However, a theoretical study \cite{fernandes12a} suggests 
that the value of $g$ becomes larger with decreasing $\delta$, so that 
$r_{\rm cr}$ is expected to be larger than in \fig{T-delta} at low $\delta$. 
The discrepancy between the present theory and 
experimental observations would even increase. 
In addition, the value of $c$ is expected to become larger for a lower $\delta$ 
because the nesting condition of the Fermi surfaces for the momenta   
$(\pi,0)$ and $(0,\pi)$ becomes better at a lower $\delta$. 
Consequently, a more realistic treatment of $g$ and $c$ would lead to  
a larger discrepancy between theory and experiment. 

\begin{figure}  [ht]
\vspace*{0cm}
\begin{center}
\includegraphics[angle=0,width=5.0cm]{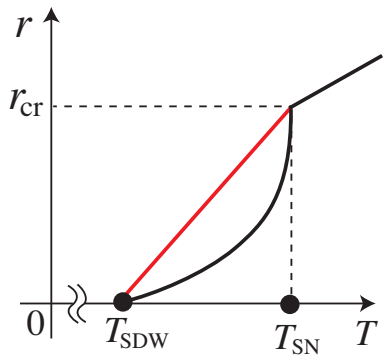} 
\end{center}
\caption{\label{r-T} (Color online) 
Temperature dependence of $r = \xi^{-2}$ ($\xi$ is the magnetic correlation
length)
in the actual system (solid curve) 
and in the linear approximation (red line). The values of $T_{\rm SN}$, $T_{\rm SDW}$
and $r_{\rm cr}$ are correctly captured in the linear approximation, although 
the whole temperature dependence of $r$ is not. The slope of the red line 
corresponds to $\alpha$ and is assumed to depend weakly on $\delta$ 
in \fig{T-delta}. 
}
\end{figure}

Below $T_{\rm SN}$ the spin nematic order parameter $\phi(T) (\geq 0)$ becomes nonzero. As 
a result $r(T)$ has the form $r^{(0)}(T) \pm \phi(T)$ where $r^{(0)}(T)$ is equal to
$r_{\rm cr} + (T-T_{\rm SN})$ in agreement with the expression \eq{r} in 
the normal state. The temperature dependence of $r$ will 
approximately be given by the  
solid curve in \fig{r-T}. The sudden drop of $r$ just below $T_{\rm SN}$ 
is due to the development of the spin nematic order parameter characterized by 
$\phi(T) \sim (T_{\rm SN}-T)^{\beta}$; $\beta=1/2$ in mean-field theory and 
$\beta=1/8$ in a two-dimensional system. The red line in \fig{r-T} is a linear 
interpolation through the points $(T,r)=(T_{\rm SN}, r_{\rm cr})$ 
and $(T_{\rm SDW}, 0)$ with the slope $\alpha$.
The absolute value of $\alpha$ simply changes the scale of temperature and thus 
is not relevant to the present discussion as long as $\alpha$ is independent of $\delta$. 

To reconcile the observed nematic phase in  terms of 
the spin nematic instability, we have to invoke a $\delta$ dependence of $\alpha$. 
In the so-called "1111" compounds LaFeAsO$_{1-x}$F$_{x}$ \cite{huang08} and 
CeFeAsO$_{1-x}$F$_{x}$ \cite{zhao08} the onset temperature of the nematic phase 
largely extends to the superconducting region even though the SDW state has already vanished. 
A similar feature is observed also for Ba(Fe$_{1-x}$Co$_x$)$_2$As$_2$ \cite{nandi10}, 
where the nematic phase is confined 
closer to the SDW phase as compared to the  "1111" compounds. 
To understand these data, $\alpha$ must be assumed to become very large 
near $x \approx 0$ 
so that $T_{\rm SDW}$ occurs much closer to $T_{\rm SN}$. It also should  
decrease substantially with increasing $x$ so that $T_{\rm SDW}$ occurs further away from 
$T_{\rm SN}$. 
This $\delta$ dependence should be so dramatic that it fully changes  
the qualitative features of the phase diagram (\fig{T-delta}) and also 
compensate the possible further discrepancy caused by a realistic treatment of $g$ and $c$. 
While the actual SDW transition changes to first order close to $x=0$ whereas 
the nematic transition is continuous there \cite{kim11a}, 
such a region is rather small and thus does not modify our major discussion. 
In typical hole-doped compounds Ba$_{1-x}$K$_{x}$Fe$_{2}$As$_{2}$ \cite{chen09,avci11} and  
isovalent doping materials Ba(Fe$_{1-x}$Ru$_{x}$)$_{2}$As$_{2}$ \cite{kim11,thaler10} 
the nematic phase occurs simultaneously with the SDW instability within experimental resolutions. 
To understand this, $\alpha$ should be assumed to be very large near $x=0$ and to remain rather large 
with increasing $x$. 

At present it is not clear whether the assumption of a $\delta$ dependence of $\alpha$ 
is justified from a microscopic point of view. To evaluate $\alpha$ 
the absolute value of $\phi(T)$ does matter as a 
function of $T$ and moreover an approximation scheme to calculate $r$ also does 
matter since $r$ becomes zero only at $T=0$ in a purely two-dimensional system. 
Because of these subtleties in determining the precise form of the spin nematic phase, 
it seems natural to expect a substantial material dependence 
of the shape of the spin nematic region in the $\delta$-$T$ plane. 
Thus one would naturally expect a phase diagram similar to \fig{T-delta} at least 
for a certain class of materials if the spin nematicity is responsible for 
the nematic phase in iron-based superconductors in general.   
At present, however, no iron-based superconductors are known to show 
characteristic features of the spin nematicity shown in \fig{T-delta} \cite{stewart11}. 
Fernandes {\it et al.} study a possible phase diagram 
of the spin nematic phase near the SDW phase \cite{fernandes12a}. 
Their calculations are done in two limits: the zero-temperature limit and 
the classical limit in the sense that only the zero Matsubara frequency is considered 
in gap equations. 
Their results at $T=0$ predict that the spin nematic and SDW instabilities occur  
simultaneously. This is consistent with our results, although they predict a first 
order transition, a  possibility which is not considered in our analysis. 
At finite temperatures they obtain in the classical limit 
figures 8 and 14 in  \cite{fernandes12a} which compare successfully with
experimental data and also present a microscopic treatment of the slope $\alpha$. 
In our approach all Matsubara frequencies are kept but $\alpha$ is considered as a phenomenological 
input. 

Our schematic phase diagram (\fig{T-delta}) 
does not apply to the orbital nematic scenario, because the orbital nematic instability 
is controlled by a fermionic bubble diagram \cite{yamase13a,yamase13b}. 
Hence it is tempting to state that in general orbital nematicity is likely responsible for 
the nematic phase observed in iron-based superconductors \cite{fisher11,stewart11}. 
However, it seems too early to reach such a conclusion. In view of the fact 
that the nematic phase occurs in general close to the SDW phase, it is important to 
clarify from a microscopic point of view why this should also hold for the orbital
nematic state.
One possible reason is given in \cite{kontani11} where Kontani {\it et al.} 
point out the important role of Aslamazov-Larkin type diagrams.  
It is interesting to explore further whether the orbital nematic scenario indeed explains 
the fact that the nematic instability occurs close to the SDW phase at $\delta=0$ and 
the nematic region extends to a larger region with decreasing $T_{\rm SDW}$ and  
higher $\delta$ 
as observed in experiments \cite{stewart11,fisher11,kasahara12}. 

The iron-based superconductor FeSe shows a structural 
phase transition from an orthorhombic to a tetragonal phase with decreasing temperature,  
but no SDW phase has been  detected \cite{mcqueen09}. 
This experimental fact naturally suggests that orbital nematicity is responsible for the 
structural phase transition. If we wish to understand the structural phase transition in FeSe 
in terms of the spin nematic order, we would invoke, for example, a large coupling strength $g$, as shown 
in the right panel in \fig{T-delta}. FeSe would then be located in the region 
$\delta_{\rm SDW} < \delta <\delta_{\rm SN}$.

\subsection{Nematic fluctuations} 
Spin nematic fluctuations can be measured directly by electronic Raman 
scattering. 
Although the computation of the Raman intensity in a microscopic model for 
electrons is beyond the scope of the present study, 
our obtained spectra can be interpreted as $B_{1g}$ Raman spectra
within the following approximation. In a microscopic calculation
the vertex $\gamma_{\vk}$ in \fig{diagram}(a) is replaced by 
a triangle diagram constructed from three electronic Green's functions. 
Such a triangle diagram depends both on momentum and frequency. 
Expanding its momentum dependence in terms of a complete set of functions
with $B_{1g}$ symmetry it is evident that only the function $\gamma({\bf k})$
contributes substantially due to the restriction of the momenta to the
neighborhood of $(\pi,0)$ and $(0,\pi)$. Concerning the frequency it is 
plausible that the triangle diagram has no resonances at low frequencies
in the range of collective nematic fluctuations. Thus the triangle diagram
may be approximated by a constant and  
we expect that our results in figures~\ref{S00-map}-\ref{S0w-w-cross} 
capture the major features of electronic Raman scattering due to spin 
nematic fluctuations in the $B_{1g}$ channel. 
Although available Raman scattering data \cite{gallais13,Zhang14,thorsmolle14} 
are interpreted in terms of 
orbital nematic fluctuations \cite{yamase13a}, they do not 
seem to exclude the spin nematic scenario. 
Not only further theoretical studies but also more detailed experimental 
data are important to determine the origin of the nematic phase observed in 
iron-based superconductors. A crucial test to identify the origin of the nematic phase 
is to measure nematic fluctuations at zero temperature by suppressing 
the superconductivity, for example, by applying a large magnetic field: 
spin nematic fluctuations appear as a soft mode upon approaching the nematic phase 
whereas orbital fluctuations form a central peak.

We finally discuss a possible role of nematic fluctuations for superconductivity. 
If the nematic phase in iron-based superconductors is associated with  spin nematicity, 
one might wonder about superconductivity mediated by spin nematic fluctuations. 
As shown in figures~\ref{S00-map} and \ref{T-delta}, spin nematic fluctuations 
are substantially suppressed at low temperatures. It is thus unlikely that 
such fluctuations are responsible for the observed superconductivity. Instead 
usual spin fluctuations provide a more 
natural scenario to understand superconductivity \cite{mazin08,kuroki08} in these sytems 
even if the spin nematic phase occurs in actual materials. 
On the other hand, if the nematic phase originates from orbital degrees of 
freedom, it has been shown that 
orbital nematic fluctuations can lead to strong coupling 
superconductivity \cite{yamase13b} in pnictides and thus provide an exotic 
mechanism for superconductivity in these systems.

\section{Conclusions}
Using a general form for the spin-fluctuation spectrum near 
the SDW phase we have studied spin nematic spectra in 
energy and momentum space. If the spin nematic interaction is attractive
and the nematic transition continuous, its transition temperature
$T_{\rm SN}$ is always higher than that to the antiferromagnetic state at 
$T_{\rm SDW}$. The spin nematic spectra are 
characterized by several generic features: 
(i) the critical magnetic correlation length $\xi_{\rm cr}$
decreases with increasing $T_{\rm SN}$ (\fig{TSN-rcr}), 
(ii) strong low-energy spin nematic fluctuations extend to a wider 
temperature region 
for a larger $T_{\rm SN}$ (\fig{S00-map}), 
(iii) the spin-nematic spectrum at $\vq = {\bf 0}$ 
exhibits a central peak as a function of $\omega$ close to the spin nematic 
instability at a finite 
temperature (figures~\ref{S0w-w} and \ref{S0w-w-cross}) whereas it shows a soft mode upon approaching 
the spin nematic instability close to zero temperature (figures~\ref{S0w-w-soft} and \ref{S0w-w-cross}), 
and (iv) there is no clear dispersive mode 
associated with spin nematic fluctuations and instead a diffusive peak is 
obtained  
around $\vq = {\bf 0}$  and $\omega=0$ (\fig{Sqw-map}). 
These general features originate from the low-energy property of 
the simple bubble diagram of spin fluctuations [see \eq{asymp}]. 
The resulting phase diagram is shown in \fig{T-delta} if $\alpha$,
the average slope of $r_{\rm cr} = \xi^{-2}_{\rm cr}$ as a function of 
temperature
between $T_{\rm SDW}$ and  $T_{\rm SN}$, can be considered as a constant.
The existing discrepancies with the experimental phase diagram may indicate that a constant
$\alpha$ is not adequate or that another scenario such as orbital nematicity
may be more appropriate.
Since a rather general form for the  spin-fluctuation spectra [see \eq{chi-1}] 
is employed in our 
study, the present theory can be applied or extended straightforwardly 
to other systems where magnetic fluctuations 
are characterized by four wavevectors with fourfold symmetry. 
Although cuprate superconductors exhibit nematicity in the magnetic excitation 
spectra \cite{hinkov08} and thus might be possible systems for spin nematic 
order, the line of onset temperatures of nematicity is nearly parallel to 
that of the incommensurate magnetic order as a function of doping in 
YBa$_{2}$Cu$_{3}$O$_{6+y}$ \cite{haug10}, 
which is at variance with \fig{T-delta}. 
Instead, the nematicity in cuprates was discussed in terms of a feedback 
effect from charge nematicity \cite{yamase09,hackl09}.

\ack
The authors thank D. V. Efremov, P. J. Hirschfeld, and T. L\"ow for stimulating 
discussions and G. Khaliullin for a critical reading of the manuscript. 
H.Y. acknowledges support by the Alexander von Humboldt Foundation 
and a Grant-in-Aid for ScientificResearch from Monkasho.     

\section*{References}
\bibliography{IOPmain.bib}%

\end{document}